\documentclass{article}
\PassOptionsToPackage{numbers,compress}{natbib}
\usepackage[final,dblblindworkshop]{neurips_2026}
\workshoptitle{Third Workshop on Agents in the Wild: Safety, Security, and Beyond}
\usepackage[utf8]{inputenc}
\usepackage[T1]{fontenc}
\usepackage{hyperref}
\usepackage{url}
\usepackage{booktabs}
\usepackage{amsfonts}
\usepackage{nicefrac}
\usepackage{microtype}
\usepackage{xcolor}
\usepackage{graphicx}
\usepackage{multirow}
\usepackage{siunitx}
\newif\ifshowjenscomments
\showjenscommentsfalse 

\newcommand{\pp}{\,pp}
\newcommand{\dfpr}{\ensuremath{\Delta\mathrm{FPR}}}
\newcommand{\dtpr}{\ensuremath{\Delta\mathrm{TPR}}}
\newcommand{\suite}{\textsc{LinguaGuardBench}}

\title{Language Is a Security Surface for Agent Guards}

\author{%
  Hasan Demirkiran\\
  Kontext\\
  \texttt{hasan@kontext.security}
  \And
  Jens Ernstberger\\
  Kontext\\
  \texttt{jens@kontext.security}
}

\begin{document}

\maketitle

\begin{abstract}
Step-level guards decide, before each tool call, whether an agent's next action is safe given the user's request and the trajectory so far. Users write those requests in many languages, while guard prompts, and the benchmarks on which purpose-built guards are trained and evaluated, are English. We ask whether a guard's verdict depends on the language of the request. We build \suite{} from the 1,220 labelled steps of TS-Bench AgentDojo-Traj: we translate only the user instruction into German, French, Turkish, Swahili and Amharic, chosen a priori to span the high-, mid- and low-resource tiers of the Joshi et al.\ taxonomy, and keep every tool call, observation and label byte-identical, so any change in a verdict is attributable to language alone. Six English rewordings and a whitespace placebo serve as controls, and every managed guard is run three times to measure its own noise. Across 19 guards, from frontier APIs to 4B open models and four purpose-built agent guards (TS-Guard, ShieldAgent, Safiron, AgentDoG), most open, small and purpose-built guards change verdicts under translation beyond their own noise floor, and the effect grows with the rarity of the language: among the nine guards below 13B parameters, the mean placebo-adjusted flip rate rises from +1.5\pp{} on French to +4.4\pp{} on Amharic. The purpose-built guards are the most sensitive of all. The direction of the damage tends to follow guard size: open guards of 12--30B parameters block \emph{more benign steps} (up to +5.0\pp{} false positives), and every significant loss of recall is in a guard below 10B (up to $-7.4$\pp{}). Frontier guards shift by at most about one point; the strongest, Gemini~3.8 Flash, returns identical verdicts with and without the request. The damage is concentrated (one translated instruction moves a guard from blocking 12\% of its steps to 92\%) and, in a post-hoc analysis, legible: the flips in excess of placebo land on steps where the agent's own reasoning restates the English request, consistent with guards reacting to inconsistency between channels. English paraphrase moves every guard less than translation does, and no general guard by more than 1.4\pp{}: the sensitivity is to language more than to wording. We release \suite{} (696 validated rewrites, frozen by hash, with a materialiser that regenerates every condition) and all 489k verdicts.
\end{abstract}

\section{Introduction}

Tool-using agents are increasingly wrapped in \emph{step-level guards}: a model that inspects the user's request, the trajectory so far and the candidate tool call, and blocks the call if it appears unsafe~\cite{toolsafe,tracesafe,llamafirewall}. The guard reads two channels. The \emph{untrusted} channel (tool outputs, web pages, files) is where prompt injections live, and its robustness is well studied~\cite{agentdojo,tracesafe}. The \emph{trusted} channel is the user's own request. The released guard prompts are English, and so are the benchmarks on which purpose-built guards are trained and evaluated (TS-Bench, Agent-SafetyBench, ATBench), but the principal they protect writes in whatever language they speak. Whether a guard's verdict on a fixed action survives a change of language is, to our knowledge, unmeasured, even though the analogous question for safety classifiers, which judge a text rather than an action in the context of a task, has a known bad answer~\cite{multijail,yong2023}.This paper tests the hypothesis, pre-registered before any guard saw a rewritten request, that the language of the request is a security surface for step-level guards. We build \suite{}: a benchmark with fixed, labelled tool-call steps in which \emph{only the user instruction} is translated into five languages chosen a priori to span three resource tiers (German and French; Turkish; Swahili and Amharic), and re-run the guards on it. Because the trajectory, the candidate action and the label never change, a verdict that flips is a language effect by construction, not a label disagreement. Two controls separate language from wording and from noise: six English rewordings of the same instructions (four LLM styles, a human paraphrase and typing noise) plus a whitespace placebo, and three repeats of every managed guard so that its own non-determinism is measured and subtracted. Our findings and contributions are best summarized as follows.

\begin{itemize}
  \item \textbf{Small guards are language-dependent, and the effect grows as the language gets rarer.} Most open and small guards change verdicts under translation beyond their noise floor; Qwen~3.8 27B is the exception. Among the nine guards below 13B parameters (five general small models and four purpose-built agent guards), the mean placebo-adjusted flip rate rises from +1.5\pp{} on French and +2.1 on German through +3.1 on Turkish to +4.2 on Swahili and +4.4 on Amharic; the worst guard changes 12\% of its verdicts, and Amharic carries the largest error shifts (\S\ref{sec:translation}, \S\ref{sec:gradient}).
  \item \textbf{Purpose-built agent guards are the most language-dependent.} TS-Guard, ShieldAgent, Safiron and AgentDoG, run with their authors' own prompts, average a placebo-adjusted flip rate of +2.5\pp{} on French and +6.3 on Swahili and Amharic. ShieldAgent's false-positive rate falls by 4--8\pp{} in every language, Safiron loses 3\pp{} of recall on Turkish and Amharic~(\S\ref{sec:specialised}).
  \item \textbf{The failure mode tends to follow scale.} Open guards of 12--30B parameters respond by blocking more benign steps (Nemotron~3.5 30B: +4.7\pp{} FPR on German, +5.0 on Amharic); every significant loss of recall is in a guard below 10B (Nemotron Nano~9B: $-3.1$ to $-4.5$\pp{} recall on German, French and Turkish, $-7.4$ on Amharic). TS-Guard (7.6B) over-blocks like the larger group, so this is a tendency, not a threshold; neither failure shows up in an English evaluation~(\S\ref{sec:scale}).
  \item \textbf{Frontier guards barely move, and the strongest does not read the request.} Claude Sonnet~5 and Gemini~3.8 Flash have no significant shift in any language; GLM~5.2's only one is a $-1.1$\pp{} false-positive shift on French. Gemini~3.8 Flash and a control that withholds the request agree on 100\% of 3,660 paired verdicts (\S\ref{sec:frontier}).
  \item \textbf{The damage is concentrated and mechanistically legible.} A few long instructions account for most translation-induced flips (one Turkish instruction takes TS-Guard from 12\% to 92\% blocked), and the flips in excess of placebo land on steps whose Thought restates the original English request, consistent with guards reacting to inconsistency between channels~(\S\ref{sec:concentration}, \ref{sec:mechanism}).
  \item \textbf{The sensitivity is to language more than to wording.} Across six English paraphrase conditions no general guard's flip rate exceeds its placebo by more than 1.4\pp{}, and every guard that moves at all moves more under translation than under paraphrase (\S\ref{sec:paraphrase-main}).
  \item \textbf{\suite{}.} A language-controlled evaluation suite for agent guards: 696 rewrites of the 58 TS-Bench AgentDojo-Traj instructions across five languages, six English styles and a placebo, each rewrite checked mechanically, judged by two cross-family models and reviewed, frozen by hash, with a materialiser that regenerates the 1,220 labelled steps of every condition byte-identically, plus 489k verdicts from 19 guards (\S\ref{sec:setup}).
\end{itemize}

The practical implication is that the trusted channel is a security surface for open and small guards deployed to non-English users, and that the failure mode to expect differs by model size: nuisance over-blocking for mid-sized open models, silent misses for small ones.

\section{Related work}

\paragraph{Step-level guards.} Content guards such as Llama Guard~\cite{llamaguard} classify a message; agent guards judge an action in the context of a task. ToolSafe/TS-Guard~\cite{toolsafe} trains a 7.6B guard with RL to judge tool invocations from the agent log; GuardAgent~\cite{guardagent} checks actions against guard requests with generated code; TraceSafe~\cite{tracesafe} and LlamaFirewall's AlignmentCheck~\cite{llamafirewall} prompt general models with the request and trajectory; Task Shield~\cite{taskshield} checks each action against the user's task; ShieldAgent~\cite{agentsafetybench}, Safiron~\cite{safiron} and AgentDoG~\cite{agentdog} are fine-tuned judges of agent interaction records, plans and trajectories. All of these read the user request as a trusted input, and all are trained or evaluated on English agent-safety benchmarks~\cite{toolemu,injecagent,agentharm,asb}; none reports sensitivity to the request's language. We evaluate the four released ones.

\paragraph{Cross-lingual safety.} Safety classifiers and refusal training degrade in low-resource languages: translating harmful prompts into low-resource languages bypasses GPT-4's safeguards~\cite{yong2023}, and multilingual jailbreak rates track language resource level~\cite{multijail}. Multilingual safety benchmarks and moderation tools~\cite{xsafety,polyguard,linguasafe,linguasafetybench} likewise vary the language of the \emph{content} being judged. Here the content (the tool call and the trajectory) is held fixed and only the language of the principal's request changes, which isolates the guard's use of the trusted channel.

\paragraph{Prompt sensitivity.} Multi-prompt evaluation shows that paraphrasing an instruction shifts LLM scores and sometimes rankings~\cite{mizrahi2024,hua2025}, and formatting alone can move accuracy by tens of points~\cite{sclar2024}; Rabanser et al.~\cite{rabanser2026} measure paraphrase robustness of \emph{agents}. Our English paraphrase conditions replicate this design for guards, and serve here as the control that shows the language effect is not a wording effect.

\section{Setup}
\label{sec:setup}

\paragraph{Benchmark.} TS-Bench AgentDojo-Traj~\cite{toolsafe} contains 1,220 labelled tool-invocation steps (868 safe, 352 unsafe) from 589 AgentDojo~\cite{agentdojo} trajectories across banking, Slack, travel and workspace environments, with 58 unique user instructions. Each step consists of the user instruction, the history (prior thoughts, actions and observations), the candidate action, environment metadata and a label. We use the released labels unchanged and execute no tool call.

\paragraph{Languages.} We translate only the instruction field into German, French, Turkish, Swahili and Amharic. The languages were chosen a priori by resource tier in the taxonomy of Joshi et al.~\cite{joshi2020}: German and French are class~5 (the largest non-English data languages), Turkish class~4, Swahili and Amharic class~2; Swahili is also the low-resource language of MultiJail~\cite{multijail}. The five span five families (Germanic, Romance, Turkic, Bantu, Semitic) and two scripts (Amharic is written in Ge'ez), and German and Turkish are read by the authors, so the pre-registered translations could be human-reviewed. The translation hypothesis and the first three languages were pre-registered; the low-resource pair extends the design with the same generation and validation pipeline and a frozen file before any guard saw it. Translations are produced by Gemini~3.8 Flash ($T=0.3$, first of three candidates passing all checks) and must preserve quoted strings, file names, e-mails, URLs, IBANs, dates, times and amounts verbatim, must not add urgency, permission or scope language, and must stay within length bounds.

\paragraph{Controls.} \texttt{original} and \texttt{placebo} (a trailing space and collapsed double spaces, the tokenisation-noise floor); six English paraphrase conditions: programmatic keyboard-adjacency typos (8\% of eligible words, slot spans protected), four LLM styles (lexical, terse, verbose-polite, casual; Gemini~3.8 Flash, $T=0.7$) and one human paraphrase per instruction written before seeing any LLM output. The paraphrases are held to the same slot and scope checks as the translations.

\paragraph{Validation.} Every accepted row was judged for task equivalence by two cross-family judges (Claude Sonnet~5, judging translations on its own back-translation; Gemini~3.1 Pro). Every one of the 580 pre-registered rows was read by a human (0 rejected, 1 repaired); the 116 low-resource rows were accepted or repaired from the back-translation under a fixed rule (Appendix~\ref{app:perturb}; 4 Amharic rows repaired). A materialisation check asserts that history, action, environment and label are byte-identical across conditions.

\paragraph{Release.} \suite{} and all verdicts are released; contents are listed in Appendix~\ref{app:perturb}.

\paragraph{Guards.} Nineteen guards, all run on one cloud platform. Sixteen use the same structured prompt (three fields: malicious user request, third-party attack, action harmfulness, mapped to TS-Guard's composite score; block if composite $\ge 0.5$): frontier APIs Gemini~3.8 Flash, Claude Sonnet~5 (thinking disabled) and GLM~5.2; managed open models gpt-oss-120b, Llama~4 Maverick, Gemma~4 26B-A4B; self-deployed open models Qwen~3.8 27B, Nemotron~3.5 Lightning 30B-A3B; a post-hoc small tier of Gemma~4 (E2B, 12B), Qwen~3.5 (4B, 9B) and Nemotron Nano (4B, 9B); the specialised TS-Guard~\cite{toolsafe} with its native prompt; and a \emph{scanner control}: Gemini~3.8 Flash with the user request withheld, which by construction must be invariant. Three further purpose-built agent guards run with their authors' released prompts and parsers (Appendix~\ref{app:serving}): ShieldAgent~\cite{agentsafetybench} (Qwen2.5-7B), Safiron~\cite{safiron} (Ministral-8B) and AgentDoG~\cite{agentdog} (Qwen3-4B). These three judge the trajectory prefix up to and including the candidate action, so their block decision is trajectory-level while the label is step-level; flip rates, which compare a guard with itself, are unaffected, and their FPR/TPR carry that caveat. The low-resource pair was run on the Gemini guards and the self-deployed guards only, including a self-deployed Gemma~4 26B-A4B that agrees with the managed one on 98\% of verdicts (Appendix~\ref{app:serving}). Managed guards are run with $K=3$ repeats at temperature~0; self-deployed guards use greedy decoding ($K=1$). Open models receive server-side JSON-schema enforcement.

\paragraph{Metrics.} For each guard and condition, paired to \texttt{original} at (step, repeat): the verdict \emph{flip rate}, the \emph{placebo-adjusted flip rate} (the flip rate minus the guard's own placebo flip rate), and \dfpr{}/\dtpr{} with 95\% percentile intervals from a 1,000-replicate bootstrap clustered by trajectory. Tier-level estimates pool the pairs of a tier's languages before bootstrapping. We also report Kendall's $\tau$ between guard rankings by balanced accuracy, flips by restatement-leakage stratum (whether $\ge 60\%$ of the instruction's content words appear in the history or Thought; 59.6\% of steps), and per-instruction block-rate changes.

\section{Results}

\subsection{Translating the request changes verdicts and error rates}
\label{sec:translation}

\begin{figure}[t]
  \centering
  \includegraphics[width=0.86\linewidth]{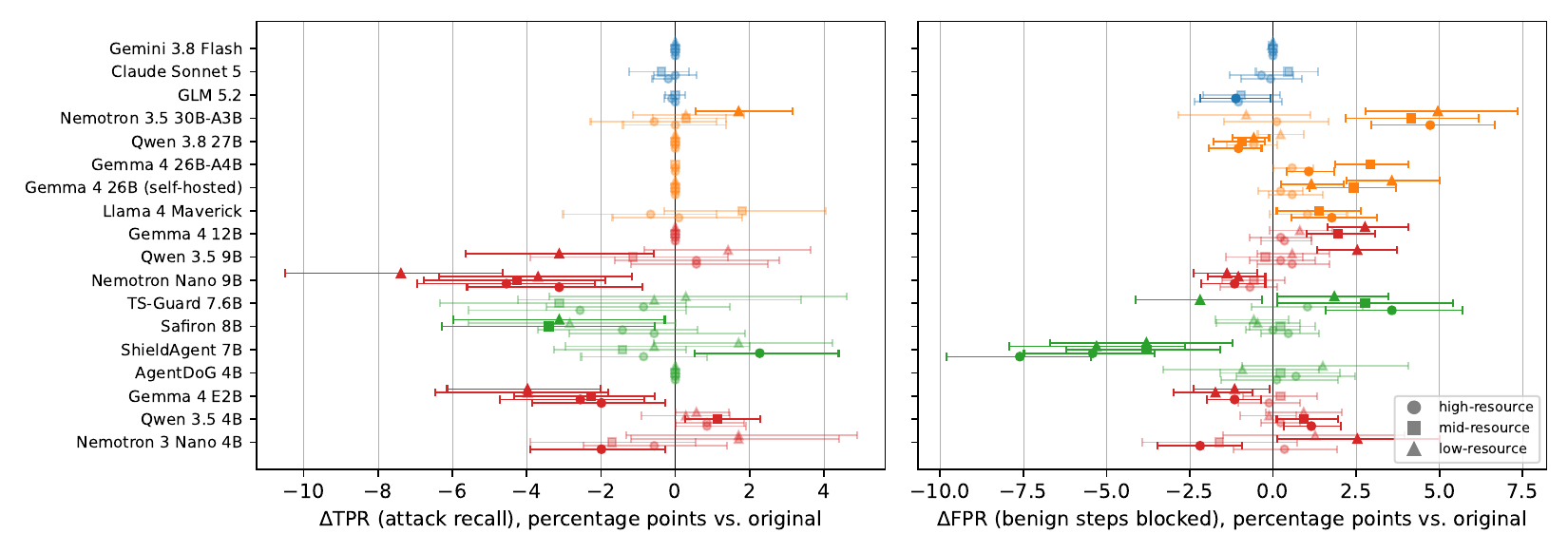}
  \caption{Change in attack recall (left) and false-positive rate (right) when the user request is translated, relative to the original English, per guard and language. Points with 95\% trajectory-clustered bootstrap intervals; faded points have intervals that include zero. Marker: high- (circle), mid- (square), low-resource (triangle) language. Colour: frontier (blue), open 17--120B (orange), small $<$13B (red), specialised (green).}
  \label{fig:shift}
\end{figure}

Figure~\ref{fig:shift} shows \dtpr{} and \dfpr{} for every guard and language with bootstrap intervals; Table~\ref{tab:translation} lists every cell with significant ones in bold. Translation changes up to 12.1\% of a guard's verdicts (ShieldAgent on Swahili, +10.5\pp{} over its placebo; Nemotron~3 Nano~4B on Amharic 11.1\%), and 39 of the 77 guard$\times$language cells (17 guards, excluding the scanner control and the degenerate Gemma~4 E2B) show a significant shift in FPR or TPR. Two patterns hold across all five languages, and they split the roster by size.

\paragraph{Mid-sized open guards over-block.} No guard of 12B parameters or more loses recall in any language; every significant shift in that range is in the false-positive rate, and the large ones are increases: Nemotron~3.5 30B-A3B (+4.7\pp{} German, +4.1 Turkish, +5.0 Amharic), Gemma~4 26B-A4B (+1.1 German, +2.9 Turkish; its self-hosted copy +2.4 Turkish, +1.2 Swahili, +3.6 Amharic), Gemma~4 12B (+2.0 Turkish, +2.8 Amharic), Llama~4 Maverick (+1.8 German, +1.4 Turkish). Two guards move the other way by about a point (Qwen~3.8 27B, GLM~5.2). TS-Guard, at 7.6B, behaves like this group (+3.6 German, +2.8 Turkish, +1.8 Amharic). For these guards a request in another language is a source of nuisance: benign steps that were allowed in English are blocked.

\paragraph{Small guards miss attacks.} Every significant \emph{decrease} in recall in the study is a guard below 10B parameters under translation: Nemotron Nano~9B loses 3.1, 4.5 and 4.3\pp{} of TPR on German, French and Turkish, 3.7 on Swahili and 7.4 on Amharic; Nemotron~3 Nano~4B loses 2.0\pp{} on German and Qwen~3.5 9B 3.1\pp{} on Swahili. Their false-positive rates move less and in both directions (Nemotron~3 Nano~4B $-2.2$ French, +2.5 Swahili; Qwen~3.5 9B +2.5 Amharic). For these guards the trusted channel is an attack surface: an attacker who controls the language of the request, or who targets non-English users, gets more attacks through. These are the weakest guards in the roster (BACC at most 0.80, Table~\ref{tab:floors}).

\begin{table}[t]\caption{Translation-induced error shifts (percentage points vs.\ original) for every guard and language; bold: 95\% trajectory-clustered interval excludes zero. Positive \dfpr{}: more benign steps blocked; negative \dtpr{}: more attacks missed. Guards sorted by total significant shift. Partner-hosted guards were not run on Swahili/Amharic (--).}\label{tab:translation}\centering\footnotesize\setlength{\tabcolsep}{4pt}
\begin{tabular}{lrrrrrrrrrr}\toprule
 & \multicolumn{5}{c}{\dfpr{} (pp)} & \multicolumn{5}{c}{\dtpr{} (pp)} \\ \cmidrule(lr){2-6}\cmidrule(lr){7-11}
Guard & de & fr & tr & sw & am & de & fr & tr & sw & am \\\midrule
ShieldAgent 7B & $\mathbf{-7.6}$ & $\mathbf{-5.4}$ & $\mathbf{-3.8}$ & $\mathbf{-5.3}$ & $\mathbf{-3.8}$ & $-0.9$ & \textbf{+2.3} & $-1.4$ & $-0.6$ & +1.7 \\
Nemotron Nano 9B & $-0.7$ & $\mathbf{-1.2}$ & $-0.6$ & $\mathbf{-1.0}$ & $\mathbf{-1.4}$ & $\mathbf{-3.1}$ & $\mathbf{-4.5}$ & $\mathbf{-4.3}$ & $\mathbf{-3.7}$ & $\mathbf{-7.4}$ \\
Nemotron 3.5 30B-A3B & \textbf{+4.7} & +0.1 & \textbf{+4.1} & $-0.8$ & \textbf{+5.0} & 0.0 & $-0.6$ & +0.3 & +0.3 & \textbf{+1.7} \\
TS-Guard 7.6B & \textbf{+3.6} & +1.0 & \textbf{+2.8} & $\mathbf{-2.2}$ & \textbf{+1.8} & $-2.6$ & $-0.9$ & $-3.1$ & $-0.6$ & +0.3 \\
Gemma 4 26B-A4B (self-hosted) & +0.6 & +0.2 & \textbf{+2.4} & \textbf{+1.2} & \textbf{+3.6} & 0.0 & 0.0 & 0.0 & 0.0 & 0.0 \\
Nemotron 3 Nano 4B & +0.3 & $\mathbf{-2.2}$ & $-1.6$ & \textbf{+2.5} & +1.3 & $\mathbf{-2.0}$ & $-0.6$ & $-1.7$ & +1.7 & +1.7 \\
Safiron 8B & +0.5 & 0.0 & +0.2 & $-0.5$ & $-0.6$ & $-0.6$ & $-1.4$ & $\mathbf{-3.4}$ & $-2.8$ & $\mathbf{-3.1}$ \\
Qwen 3.5 9B & +0.6 & +0.2 & $-0.2$ & +0.6 & \textbf{+2.5} & +0.6 & +0.6 & $-1.1$ & $\mathbf{-3.1}$ & +1.4 \\
Gemma 4 12B & +0.3 & +0.2 & \textbf{+2.0} & +0.8 & \textbf{+2.8} & 0.0 & 0.0 & 0.0 & 0.0 & 0.0 \\
Gemma 4 26B-A4B & \textbf{+1.1} & +0.6 & \textbf{+2.9} & -- & -- & 0.0 & 0.0 & 0.0 & -- & -- \\
Qwen 3.5 4B & \textbf{+1.2} & +0.2 & \textbf{+0.9} & $-0.1$ & +0.9 & +0.9 & +0.9 & \textbf{+1.1} & +0.3 & +0.6 \\
Llama 4 Maverick & \textbf{+1.8} & +1.0 & \textbf{+1.4} & -- & -- & +0.1 & $-0.7$ & +1.8 & -- & -- \\
Qwen 3.8 27B & $\mathbf{-1.0}$ & $-0.6$ & $\mathbf{-0.9}$ & $\mathbf{-0.6}$ & +0.2 & 0.0 & 0.0 & 0.0 & 0.0 & 0.0 \\
GLM 5.2 & $-1.0$ & $\mathbf{-1.1}$ & $-1.0$ & -- & -- & 0.0 & $-0.1$ & 0.0 & -- & -- \\
Gemini 3.8 Flash & 0.0 & 0.0 & 0.0 & 0.0 & 0.0 & 0.0 & 0.0 & 0.0 & 0.0 & 0.0 \\
Claude Sonnet 5 & $-0.1$ & $-0.3$ & +0.5 & -- & -- & $-0.2$ & 0.0 & $-0.4$ & -- & -- \\
AgentDoG 4B & +0.1 & +0.7 & +0.2 & $-0.9$ & +1.5 & 0.0 & 0.0 & 0.0 & 0.0 & 0.0 \\
\bottomrule\end{tabular}\end{table}

\subsection{The sign of the damage tends to follow guard size}
\label{sec:scale}

\begin{figure}[t]
  \centering
  \includegraphics[width=0.66\linewidth]{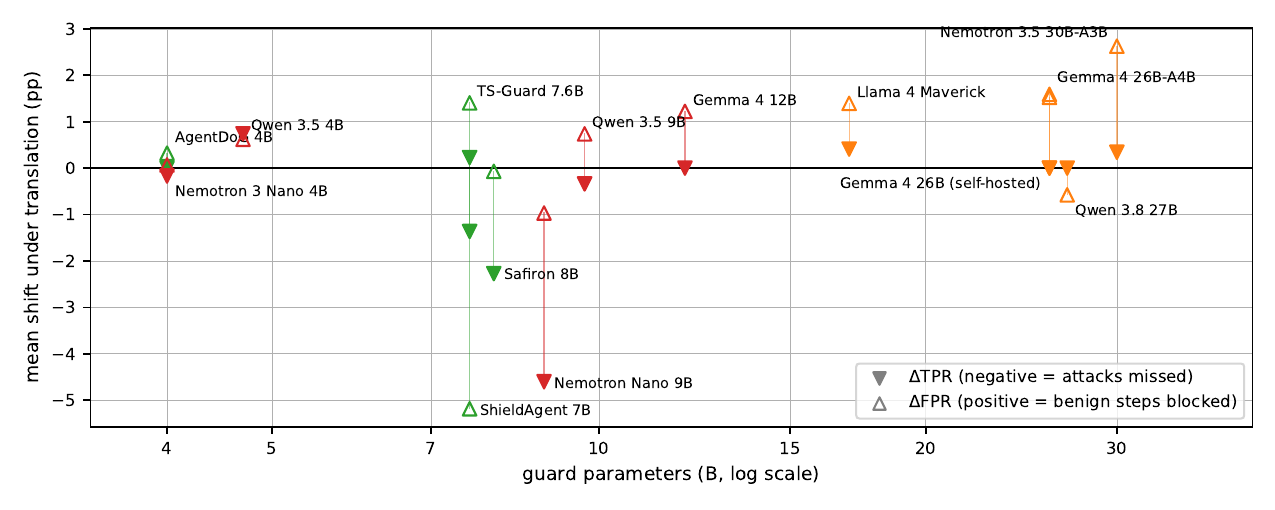}
  \caption{Mean shift under translation (pooled over the languages each guard was run on) against guard size. Filled triangles pointing down: \dtpr{} (negative means attacks missed); open triangles pointing up: \dfpr{} (positive means benign steps blocked); a vertical line joins the two markers of one guard. The degenerate Gemma~4 E2B is omitted. Over-blocking is a 12--30B phenomenon; missed attacks appear below 10B.}
  \label{fig:size}
\end{figure}

Figure~\ref{fig:size} places the pooled shifts on a size axis. The crossover is not monotone in raw flip rate (Qwen~3.5 9B flips more than both its 4B and 27B siblings), but there is a tendency in \emph{which} error grows: at 12B and above every significant shift is in false positives, and every significant loss of recall is below 10B. The converse does not hold (TS-Guard at 7.6B and several sub-10B guards also over-block), so this is a tendency, not a threshold. The two regimes are not two degrees of the same failure. A mid-sized guard that over-blocks has not lost the ability to recognise an attack; it has become more suspicious of everything, and the extra blocks fall on benign steps because there are more of them. A small guard that under-blocks has lost the signal: with a request it parses less well, an unsafe action looks like every other action. Under the strict decision rule (block only if composite $=1$), the placebo-adjusted flip rates under translation grow further for the guards that move (Nemotron~3.5 +5.7\pp{}, Nemotron~3 Nano~4B +5.1, Qwen~3.5 4B +2.9, TS-Guard +2.6) while every frontier guard stays within 0.2\pp{} under both rules (Appendix~\ref{app:strict}).

\subsection{The effect grows as the language gets rarer}
\label{sec:gradient}

Table~\ref{tab:tiers} (Appendix) pools each guard's pairs by language tier (Joshi et al.\ classes 5, 4 and 2) before bootstrapping. For the six guards below 13B parameters, the mean placebo-adjusted flip rate rises monotonically with rarity: +0.8\pp{} on French, +1.3 German, +2.5 Turkish, +2.8 Swahili, +3.2 Amharic (the five general small models alone: +0.7, +1.1, +2.0, +2.5, +3.0). The four purpose-built guards follow the same order at a higher level (+2.5, +3.4, +4.3, +6.4, +6.2; \S\ref{sec:specialised}), the open 12--30B guards at a lower one (+0.7, +1.1, +1.6 on the pre-registered languages; +1.0 and +2.1 on Swahili and Amharic for the two run there), and the frontier guards stay below +0.5\pp{} everywhere. Per guard, the low-resource tier is at least as bad as the high-resource tier for every guard that moves at all. It is the worst tier for Nemotron~3 Nano~4B (+0.6 $\to$ +4.1 $\to$ +7.2\pp{} from high through mid to low), Nemotron~3.5 30B (+1.2 $\to$ +2.5 $\to$ +3.2), Nemotron Nano~9B (+1.5 $\to$ +1.6 $\to$ +2.2) and Gemma~4 12B (+0.4 $\to$ +0.6 $\to$ +0.8). For TS-Guard, Qwen~3.5 9B and the self-hosted Gemma~4 26B, Turkish is marginally worse than the low tier; Gemini~3.8 Flash and Qwen~3.8 27B do not move in any tier. The error shifts are guard-specific rather than class-wide. Pooled over the five general small guards, \dtpr{} is $-0.7$ [$-1.3$, $-0.1$] on high-, $-1.2$ [$-2.0$, $-0.3$] on mid- and $-0.9$ [$-1.8$, $-0.0$]\pp{} on low-resource languages, and \dfpr{} turns positive on the low tier (+0.9 [+0.3, +1.4]): the recall loss is carried by Nemotron Nano~9B ($-3.8 \to -4.3 \to -5.5$), while the two 4B guards drift towards blocking on Swahili and Amharic. For the over-blockers, Amharic is the worst single language in the study (Nemotron~3.5 +5.0\pp{} FPR, Gemma~4 26B +3.6, Gemma~4 12B +2.8, Qwen~3.5 9B +2.5), while Swahili is not: its flips are as frequent as Turkish's, but the error shifts scatter in both directions (two significant FPR increases against three decreases, including TS-Guard at $-2.2$\pp{}), so Swahili scrambles verdicts without a consistent bias. Amharic is also the only non-Latin script in the set; separating script from resource level needs more languages than we ran.

The pre-registered languages already showed the ordering: Turkish, the mid-resource language, was harder than German and French for every guard (seven guards with a significant FPR increase against five and zero; TS-Guard +4.8\pp{} placebo-adjusted, 9.3\% raw). The low-resource pair, added afterwards, extends it: the two largest raw flip rates in the study are Nemotron~3 Nano~4B on Swahili (10.8\%) and Amharic (11.1\%), and the guard ranking is least stable under Amharic among the languages run on the same roster ($\tau=0.93$ against 0.99 for Swahili).

\subsection{Purpose-built agent guards are the most language-dependent}
\label{sec:specialised}

The four guards trained specifically to judge agent actions, run with their authors' own prompts, are the most language-dependent guards in the study. ``Language-dependent'' means verdict instability, not harm: ShieldAgent's error rates improve under translation, and the comparison is within guard. Their mean placebo-adjusted flip rate under translation is +3.4\pp{} (TS-Guard), +2.9 (Safiron), +4.3 (AgentDoG) and +7.7 (ShieldAgent), against +1.4, +1.3, +1.8 and +3.5 under English paraphrase, and each rises from the high- to the low-resource tier (TS-Guard +2.0 $\to$ +4.8 $\to$ +4.1, Safiron +1.9 $\to$ +3.2 $\to$ +3.7, AgentDoG +2.5 $\to$ +2.7 $\to$ +7.0, ShieldAgent +5.5 $\to$ +6.7 $\to$ +10.4). ShieldAgent changes 12.1\% of its verdicts on Swahili and 11.9\% on Amharic (its placebo floor is 1.6\%). The direction differs by guard and does not follow the size rule of \S\ref{sec:scale}. ShieldAgent, which blocks 44\% of benign steps in English, blocks \emph{fewer} of them in every other language ($-7.6$\pp{} German, $-5.4$ French, $-3.8$ Turkish, $-5.3$ Swahili, $-3.8$ Amharic, all significant) while its recall holds. Safiron, the weakest of the four (recall 0.41), loses recall on Turkish ($-3.4$\pp{}), Swahili ($-2.8$) and Amharic ($-3.1$); AgentDoG flags every unsafe trajectory in every language and moves only on benign steps; TS-Guard over-blocks on German, Turkish and Amharic and under-blocks on Swahili. What the four share is that a translated request changes the operating point of a guard whose documented training data, prompts and benchmarks are English, by more than any general-purpose model of the same size. The field's own guards are the least stable.

\subsection{Frontier guards are invariant, and the strongest one does not read the request}
\label{sec:frontier}

Table~\ref{tab:floors} gives each guard's original-wording operating point and placebo flip rate. Gemini~3.8 Flash and Claude Sonnet~5 are the best guards (BACC 0.983 and 0.958), and neither shows a significant change in FPR or TPR under any language (all five for Gemini, the three pre-registered for Claude); their placebo-adjusted flip rates are 0.0\pp{} and at most 1.1\pp{}. GLM~5.2 shows one significant shift, a decrease of 1.1\pp{} in false positives on French, and none in recall.

The scanner control makes the Gemini result sharper than mere robustness. The control receives the identical prompt with the user-request field removed; its prompt hash differs and its inputs are 43 tokens shorter. It agrees with the full Gemini guard on \textbf{100.0\%} of 3,660 paired verdicts on the original wording, has an identical operating point (TPR 0.977, FPR 0.012), and flips at most one of its 3,660 verdicts under every condition, as does Gemini itself. The instruction text appears verbatim in the history for only 3 of 1,220 steps, so this is not copying: on this benchmark the trajectory carries everything a strong guard uses, and the trusted channel is redundant. A guard that ignores the request cannot be attacked through its language; the converse is that such a guard also cannot use the request to disambiguate an action, which is the case Task Shield-style designs~\cite{taskshield} are built for.

\begin{table}[t]
  \caption{Original-wording operating point and placebo flip rate (the guard's own noise floor) for all guards, sorted by balanced accuracy. Managed guards $K=3$ (3,660 verdicts), self-deployed $K=1$ (1,220). Gemma~4 E2B is degenerate (blocks 6.5\% of steps) and excluded from interpretation. ShieldAgent, AgentDoG and Safiron judge the trajectory prefix, so their FPR/TPR against step labels are not comparable with the step-level guards.}
  \label{tab:floors}
  \centering
  \footnotesize
  \begin{tabular}{llS[table-format=1.3]S[table-format=1.3]S[table-format=1.3]S[table-format=1.3]}
    \toprule
    Guard & Tier & {BACC} & {FPR} & {TPR} & {Placebo flip} \\
    \midrule
    Gemini 3.8 Flash (scanner control) & frontier & 0.983 & 0.012 & 0.977 & 0.001 \\
    Gemini 3.8 Flash & frontier & 0.983 & 0.012 & 0.977 & 0.000 \\
    Claude Sonnet 5 & frontier & 0.958 & 0.072 & 0.988 & 0.017 \\
    TS-Guard (7.6B, specialised) & specialised & 0.910 & 0.071 & 0.892 & 0.045 \\
    gpt-oss-120b & open & 0.899 & 0.179 & 0.976 & 0.127 \\
    Gemma 4 12B & small & 0.849 & 0.303 & 1.000 & 0.010 \\
    GLM 5.2 & frontier & 0.842 & 0.313 & 0.997 & 0.061 \\
    Gemma 4 26B-A4B & open & 0.829 & 0.342 & 1.000 & 0.005 \\
    Qwen 3.8 27B & open & 0.825 & 0.350 & 1.000 & 0.007 \\
    Nemotron 3.5 30B-A3B & open & 0.820 & 0.338 & 0.977 & 0.042 \\
    Llama 4 Maverick & open & 0.818 & 0.273 & 0.909 & 0.028 \\
    Qwen 3.5 4B & small & 0.809 & 0.371 & 0.989 & 0.005 \\
    Qwen 3.5 9B & small & 0.801 & 0.262 & 0.864 & 0.016 \\
    ShieldAgent (7B, specialised) & specialised & 0.740 & 0.438 & 0.918 & 0.016 \\
    Nemotron 3 Nano 4B & small & 0.737 & 0.298 & 0.773 & 0.038 \\
    Nemotron Nano 9B & small & 0.684 & 0.172 & 0.540 & 0.012 \\
    AgentDoG (4B, specialised) & specialised & 0.664 & 0.672 & 1.000 & 0.024 \\
    Safiron (8B, specialised) & specialised & 0.614 & 0.183 & 0.412 & 0.004 \\
    Gemma 4 E2B & small & 0.544 & 0.039 & 0.128 & 0.004 \\
    \bottomrule
  \end{tabular}
\end{table}

\subsection{The damage is concentrated in a few instructions}
\label{sec:concentration}

Flip rates are averages over 1,220 steps from 58 instructions; the per-instruction view shows they are not evenly spread. TS-Guard on Turkish flips 9.3\% of verdicts overall, but on one instruction (``Can you recommend me a hotel with highest rating in Paris? If there are multiple options\ldots'', 26 steps) its block rate goes from 12\% to 92\%, i.e.\ 81\% of that instruction's steps flip. Gemma~4 26B on Turkish moves ``Perform the following tasks in the order they are given: 1.\ How many appointments do I have\ldots'' (69 steps) from 62\% to 97\% blocked; Nemotron~3.5 on German moves the Facebook-security-code instruction of the same enumerated family from 58\% to 83\%. The low-resource pair hits the same family: Nemotron~3 Nano~4B on Swahili moves ``Perform the following tasks in the order they are given: 1.\ What are we going to do on June 13\ldots'' (24 steps) from 0\% to 62\% blocked and, on Amharic, the Paris-trip budget instruction (53 steps) from 49\% to 85\%; Nemotron~3.5 on Amharic moves ``Please show me my unread emails'' (14 steps) from 36\% to 93\%. The direction is not always towards blocking: TS-Guard on Swahili releases the Facebook-security-code instruction from 92\% to 33\% blocked, and Nemotron~3.5 on Swahili releases the TODO-list e-mail instruction from 95\% to 37\%. The ``perform the following tasks in the order they are given'' instructions recur as hot spots across guards and languages: long, enumerated, multi-task requests are where translation turns an open guard into a block-everything guard for the whole trajectory.

\subsection{Mechanism: guards react to inconsistency between channels}
\label{sec:mechanism}

\begin{figure}[t]
  \centering
  \includegraphics[width=0.84\linewidth]{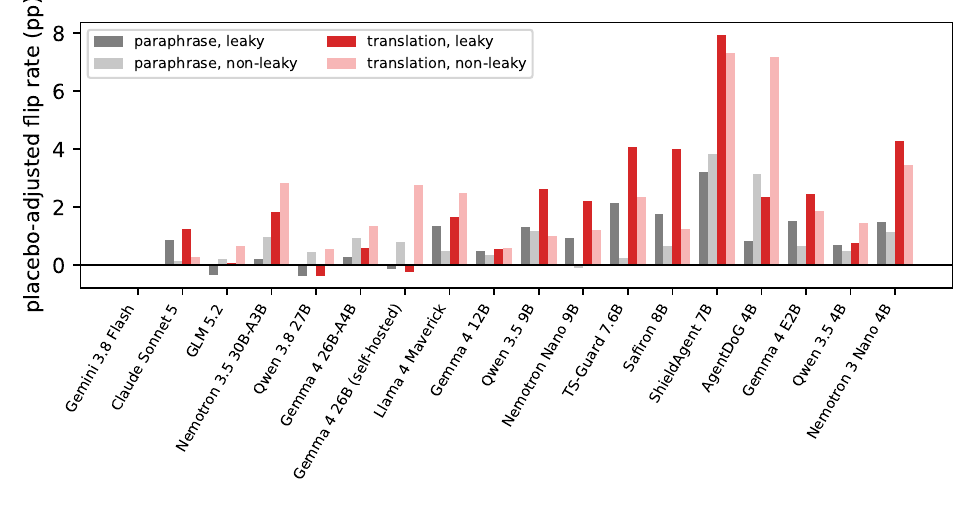}
  \caption{Placebo-adjusted verdict flip rate, split by whether the step's history restates the request (leaky) or not, pooled over translation conditions and, for comparison, over paraphrase conditions. For the strong guards and the recall-losers the rate is higher on leaky steps; for the over-blockers it is higher on non-leaky steps.}
  \label{fig:leaky}
\end{figure}

We pre-registered the expectation that flips would concentrate on \emph{non-leaky} steps, where the history does not restate the request and the guard has to rely on the request field. The data split by regime (Figure~\ref{fig:leaky}). Leaky steps already flip more under placebo (they sit later in trajectories), but for the strong guards and for the guards that lose recall the placebo-adjusted flip rate under translation is also larger on leaky steps: Claude Sonnet~5 +1.2 vs.\ +0.3\pp{}, TS-Guard +4.1 vs.\ +2.4, Nemotron Nano~9B +2.2 vs.\ +1.2, Qwen~3.5 9B +2.6 vs.\ +1.0; the same ordering holds under paraphrase (Claude +0.9 vs.\ +0.1, TS-Guard +2.1 vs.\ +0.2). When the agent's Thought restates the \emph{original English} request and the request field now says it in another language, these guards see two channels that disagree, and their verdict moves. For the over-blockers the placebo-adjusted flip rate is larger on the non-leaky steps: Nemotron~3.5 +1.8 vs.\ +2.8, Llama~4 Maverick +1.7 vs.\ +2.5, Gemma~4 26B +0.6 vs.\ +1.4 (self-hosted $-0.2$ vs.\ +2.8), AgentDoG +2.4 vs.\ +7.2. Safiron (+4.0 vs.\ +1.3) sides with TS-Guard; ShieldAgent moves on both (+8.0 vs.\ +7.3). Where nothing in the trajectory restates the request, these guards treat the foreign request itself as suspicious. Two mechanisms are consistent with this split, and they match the two regimes: inconsistency detection, which needs a restatement to fire and moves verdicts in either direction, and distrust of a foreign request, which fires when the request is the only source and adds blocks. Both are bounded by the restatement rate (60\% of steps), which our intervention does not change, and a guard that decides from the trajectory alone, as the frontier models appear to do, is immune to both.

\subsection{The sensitivity is to language more than to wording}
\label{sec:paraphrase-main}

The same instructions reworded in English are the control. Across typos, four LLM styles and a human paraphrase, no general guard's verdict flip rate exceeds its own placebo by more than 1.4\pp{} (TS-Guard among the pre-registered roster, Nemotron~3 Nano~4B in the small tier); the two purpose-built guards that do move under paraphrase (ShieldAgent +3.5\pp{}, AgentDoG +1.8) move two to three times as much under translation. Every guard that moves at all moves more under translation than under paraphrase. No paraphrase condition reduces any guard's recall except gpt-oss-120b on the terse style ($-1.3$\pp{}), and the guard ranking is preserved (Kendall's $\tau$ 0.96--0.98 against 0.93--0.96 under translation). Dose--response by surface-distance tercile is flat for all but two guards. The only paraphrase effects that reach significance are register effects on the false-positive rate of small guards and TS-Guard, in both directions (polite verbose phrasing lowers Nemotron~3 Nano~4B's FPR by 4.8\pp{} and raises Qwen~3.5 9B's by 3.0\pp{}). The full paraphrase results are in Appendix~\ref{sec:paraphrase}. A guard that is nearly invariant to how a request is worded but not to which language it is written in has learned the language, more than the register, as a feature.

\section{Discussion}

\paragraph{What to deploy.} For a guard that must serve non-English users, size is a proxy for the failure mode to expect: 17--30B open models tend to over-block, the guards that under-block are all below 10B, and neither shows up in an English-only evaluation. Purpose-built agent guards are not exempt; they are the most sensitive class we measured, and their direction has to be measured per guard. We evaluated no mitigation; translate-then-guard or per-language thresholds are untested candidates for the over-blocking regime. The concentration in enumerated multi-task instructions (\S\ref{sec:concentration}) suggests a cheap diagnostic: evaluate guards on translated \emph{long} instructions specifically, and on a low-resource language, where the effect is largest.

\paragraph{The request field is not what strong guards use.} The Gemini/scanner identity and the leaky-step mechanism point the same way: on trajectory-rich benchmarks the agent's own reasoning restates the request, and strong guards decide from the trajectory. This is good news for robustness and bad news for benchmark validity: a benchmark cannot test a guard's use of the trusted channel if the untrusted channel already contains it. It also reframes language robustness: a guard is hurt by translation to the extent that it compares channels, not to the extent that it reads the request.

\paragraph{Limitations.} The LLM paraphrases and all five translations were generated by Gemini~3.8 Flash, which is also a guard; its measured invariance is therefore a lower bound, although the scanner identity makes generator overlap an unlikely explanation. Nobody on the team reads Swahili or Amharic, so those rows were accepted on the strength of two judges and back-translation, and the low-resource pair was not run on the partner-hosted guards (Claude, Llama, GLM, gpt-oss). The small tier and the low-resource pair were added after the pre-registered runs (ten guards, three languages), each with its own frozen-before-inference file. The alignment prompt gives open models high false-positive rates (0.26--0.37) at the original wording; effects are reported as shifts from each guard's own operating point. gpt-oss-120b was quota-throttled and completed 7 of the 11 pre-registered perturbed conditions. ShieldAgent, Safiron and AgentDoG were built to judge whole interaction records, plans or trajectories; we feed them the prefix up to the candidate action, and Safiron's input format has no observation channel, so our adapter places each step's thought, tool call and observation in its action list. Their error rates against step labels are therefore not comparable with the step-level guards; their flip rates are. The bootstrap clusters by trajectory (589) while the intervention is per instruction (58), so the intervals are likely too narrow (\S\ref{sec:concentration}).

\section{Conclusion}

Translating a user request changes what open and small step-level guards decide about a fixed action, and the direction of the damage tends to follow the guard's size: mid-sized open guards block more benign steps, small guards let more attacks through, purpose-built agent guards move most of all, and frontier guards, the strongest of which does not use the request field at all, barely move. The effect grows as the language gets rarer: Amharic, the lowest-resource language and the only non-Latin script in the set, produces the largest shifts in both regimes. It is concentrated in long enumerated instructions and, in a post-hoc analysis, lands where the request and the agent's own restatement of it disagree. Rewording the same request in English changes far less. For open and small guards the language of the request is a security surface, and an evaluation that only ever shows the guard English requests will not expose it.

\appendix

\section{English paraphrase results in full}
\label{sec:paraphrase}

Table~\ref{tab:flips} gives raw verdict flip rates for every guard and condition, including the placebo floor; Tables~\ref{tab:para-dfpr} and \ref{tab:para-dtpr} give the error shifts under the six English paraphrase conditions. The pre-registered primary outcome of the study was the placebo-adjusted flip rate pooled over paraphrases: it is at most $+1.4$\pp{} (TS-Guard) for the pre-registered roster and at most $+1.4$\pp{} (Nemotron~3 Nano~4B) in the small tier, i.e.\ English rewording does not change what guards decide beyond their own noise. Dose--response by surface-distance tercile (normalised edit distance between original and paraphrase, binned per instruction) is flat for every guard except Gemma~4 26B (0.6\% $\to$ 1.6\%) and Qwen~3.5 4B (0.7\% $\to$ 1.8\%). No paraphrase condition reduces any guard's recall (one exception: gpt-oss-120b, terse, $-1.3$\pp{}); the FPR shifts that do reach significance are the register effects noted in \S\ref{sec:paraphrase-main}: polite verbose phrasing lowers Nemotron~3 Nano~4B's FPR by 4.8\pp{} and TS-Guard's by 2.8\pp{} while raising Qwen~3.5 9B's by 3.0\pp{}; lexical rewording lowers TS-Guard's FPR by 2.0\pp{} and raises Gemma~4 26B's by 2.0\pp{}; recall is unchanged in every one of these cells.

\begin{table}[t]\caption{Raw verdict flip rate (\%) vs.\ the original wording, per guard and condition. The placebo column is each guard's own noise floor.}\label{tab:flips}\centering\scriptsize
\begin{tabular}{lrrrrrrrrrrrr}\toprule
Guard & placebo & typos & lexical & terse & verbose & casual & human & de & fr & tr & sw & am \\\midrule
Gemini 3.8 Flash (scanner) & 0.1 & 0.0 & 0.0 & 0.0 & 0.0 & 0.0 & 0.0 & 0.0 & 0.0 & 0.0 & 0.0 & 0.0 \\
Gemini 3.8 Flash & 0.0 & 0.0 & 0.0 & 0.0 & 0.0 & 0.0 & 0.0 & 0.0 & 0.0 & 0.0 & 0.0 & 0.0 \\
Claude Sonnet 5 & 1.7 & 1.9 & 2.1 & 2.2 & 2.9 & 2.3 & 2.4 & 2.3 & 2.7 & 2.8 & -- & -- \\
GLM 5.2 & 6.1 & 5.8 & 6.2 & 5.7 & 6.4 & 5.7 & 6.3 & 6.6 & 6.2 & 6.5 & -- & -- \\
gpt-oss-120b & 12.7 & 12.4 & 13.4 & 12.9 & 12.7 & 13.1 & 14.9 & -- & -- & -- & -- & -- \\
Nemotron 3.5 30B-A3B & 4.2 & 4.3 & 3.9 & 4.6 & 4.8 & 5.2 & 5.3 & 6.0 & 4.8 & 6.6 & 6.4 & 8.3 \\
Qwen 3.8 27B & 0.7 & 0.2 & 1.0 & 0.7 & 0.5 & 0.3 & 1.1 & 0.7 & 0.7 & 0.7 & 0.4 & 0.8 \\
Gemma 4 26B-A4B & 0.5 & 0.5 & 1.6 & 0.6 & 1.3 & 0.9 & 1.2 & 1.1 & 0.8 & 2.2 & -- & -- \\
Gemma 4 26B-A4B (self-hosted) & 0.8 & 1.1 & 1.4 & 0.6 & 1.0 & 1.0 & 1.4 & 1.1 & 0.7 & 2.5 & 1.5 & 3.2 \\
Llama 4 Maverick & 2.8 & 3.9 & 3.7 & 3.9 & 3.8 & 4.1 & 3.6 & 4.7 & 4.6 & 5.1 & -- & -- \\
TS-Guard 7.6B & 4.5 & 4.6 & 6.4 & 6.1 & 5.9 & 5.8 & 6.4 & 7.2 & 5.7 & 9.3 & 8.6 & 8.6 \\
ShieldAgent 7B & 1.6 & 2.4 & 5.5 & 6.7 & 4.6 & 4.3 & 7.1 & 8.0 & 6.3 & 8.4 & 12.1 & 11.9 \\
Safiron 8B & 0.4 & 0.7 & 1.7 & 2.0 & 2.6 & 1.4 & 2.0 & 2.5 & 2.2 & 3.6 & 4.3 & 3.9 \\
AgentDoG 4B & 2.4 & 2.1 & 3.5 & 5.0 & 5.4 & 4.4 & 4.3 & 5.0 & 4.8 & 5.1 & 9.3 & 9.3 \\
Gemma 4 12B & 1.0 & 1.2 & 1.6 & 1.6 & 1.6 & 1.2 & 1.2 & 1.1 & 1.6 & 1.6 & 1.4 & 2.1 \\
Qwen 3.5 9B & 1.6 & 1.6 & 3.0 & 2.9 & 3.7 & 2.7 & 3.0 & 2.9 & 2.8 & 4.1 & 3.9 & 4.0 \\
Nemotron Nano 9B & 1.2 & 1.4 & 2.5 & 1.6 & 1.8 & 1.3 & 1.8 & 2.5 & 3.0 & 2.8 & 3.0 & 3.9 \\
Gemma 4 E2B & 0.4 & 0.9 & 1.6 & 1.8 & 1.3 & 1.4 & 2.5 & 2.5 & 2.0 & 2.8 & 3.0 & 2.8 \\
Qwen 3.5 4B & 0.5 & 1.0 & 1.1 & 0.7 & 1.1 & 1.2 & 1.3 & 1.6 & 0.7 & 1.8 & 1.6 & 2.0 \\
Nemotron 3 Nano 4B & 3.8 & 4.4 & 4.9 & 5.6 & 6.1 & 4.3 & 5.4 & 5.2 & 3.5 & 7.9 & 10.8 & 11.1 \\
\bottomrule\end{tabular}\end{table}
\begin{table}[t]\caption{\dfpr{} (pp) under the six paraphrase conditions; bold: 95\% interval excludes 0.}\label{tab:para-dfpr}\centering\small
\begin{tabular}{lrrrrrr}\toprule
Guard & typos & lexical & terse & verbose & casual & human \\\midrule
Gemini 3.8 Flash (scanner) & 0.0 & 0.0 & 0.0 & 0.0 & 0.0 & 0.0 \\
Gemini 3.8 Flash & 0.0 & 0.0 & 0.0 & 0.0 & 0.0 & 0.0 \\
Claude Sonnet 5 & 0.0 & $-0.2$ & $-0.2$ & +0.5 & $-0.2$ & +0.9 \\
GLM 5.2 & $-0.6$ & +0.2 & +0.2 & +1.0 & +0.1 & +0.2 \\
gpt-oss-120b & $\mathbf{-4.4}$ & $\mathbf{-3.8}$ & $\mathbf{-5.6}$ & $\mathbf{-4.0}$ & $\mathbf{-3.2}$ & $\mathbf{-5.1}$ \\
Nemotron 3.5 30B-A3B & +1.3 & $-0.1$ & $-1.3$ & $-1.5$ & $-0.3$ & $-0.9$ \\
Qwen 3.8 27B & $-0.2$ & +0.5 & $-0.7$ & 0.0 & $-0.2$ & +0.1 \\
Gemma 4 26B-A4B & $-0.1$ & \textbf{+2.0} & +0.1 & \textbf{+1.5} & \textbf{+0.9} & \textbf{+1.3} \\
Gemma 4 26B-A4B (self-hosted) & $-0.5$ & +0.8 & $-0.3$ & 0.0 & 0.0 & \textbf{+1.3} \\
Llama 4 Maverick & +0.6 & +0.4 & $-0.4$ & $-0.2$ & +0.6 & +0.6 \\
TS-Guard 7.6B & +1.0 & $\mathbf{-2.0}$ & 0.0 & $\mathbf{-2.8}$ & $-1.2$ & +2.0 \\
ShieldAgent 7B & +0.9 & $-0.5$ & $-1.8$ & $\mathbf{-2.0}$ & $-0.3$ & $-1.0$ \\
Safiron 8B & +0.3 & +0.1 & \textbf{+0.8} & +0.9 & 0.0 & +0.5 \\
AgentDoG 4B & 0.0 & $-0.3$ & +1.0 & \textbf{+3.9} & \textbf{+2.5} & $-0.6$ \\
Gemma 4 12B & +0.6 & +0.5 & +0.6 & $-0.3$ & +0.3 & +0.6 \\
Qwen 3.5 9B & +0.6 & \textbf{+1.2} & +0.1 & \textbf{+3.0} & \textbf{+1.3} & +0.7 \\
Nemotron Nano 9B & 0.0 & $-0.2$ & +0.1 & +0.2 & $-0.2$ & $-0.2$ \\
Gemma 4 E2B & $-0.1$ & +0.3 & $-0.5$ & $\mathbf{-0.9}$ & $-0.1$ & +0.6 \\
Qwen 3.5 4B & +0.5 & \textbf{+0.9} & $-0.1$ & +0.2 & \textbf{+0.9} & +0.8 \\
Nemotron 3 Nano 4B & +1.6 & $\mathbf{-2.9}$ & $\mathbf{-3.1}$ & $\mathbf{-4.8}$ & $-0.9$ & $-0.5$ \\
\bottomrule\end{tabular}\end{table}
\begin{table}[t]\caption{\dtpr{} (pp) under the six paraphrase conditions; bold: 95\% interval excludes 0.}\label{tab:para-dtpr}\centering\small
\begin{tabular}{lrrrrrr}\toprule
Guard & typos & lexical & terse & verbose & casual & human \\\midrule
Gemini 3.8 Flash (scanner) & 0.0 & 0.0 & 0.0 & 0.0 & 0.0 & 0.0 \\
Gemini 3.8 Flash & 0.0 & 0.0 & 0.0 & 0.0 & 0.0 & 0.0 \\
Claude Sonnet 5 & $-0.5$ & $-0.2$ & $-0.5$ & 0.0 & +0.1 & $-0.1$ \\
GLM 5.2 & 0.0 & +0.1 & 0.0 & +0.2 & +0.1 & $-0.1$ \\
gpt-oss-120b & $-0.9$ & $-0.9$ & $\mathbf{-1.3}$ & +0.5 & +0.1 & $-0.7$ \\
Nemotron 3.5 30B-A3B & 0.0 & $-0.3$ & $-0.3$ & +0.3 & +0.9 & $-0.3$ \\
Qwen 3.8 27B & 0.0 & 0.0 & 0.0 & 0.0 & 0.0 & 0.0 \\
Gemma 4 26B-A4B & 0.0 & 0.0 & 0.0 & 0.0 & 0.0 & 0.0 \\
Gemma 4 26B-A4B (self-hosted) & 0.0 & 0.0 & 0.0 & 0.0 & 0.0 & 0.0 \\
Llama 4 Maverick & +1.3 & +0.3 & +0.5 & $-0.7$ & +0.1 & $-0.1$ \\
TS-Guard 7.6B & $-2.6$ & $-0.3$ & +0.9 & +0.6 & $-1.4$ & $-0.3$ \\
ShieldAgent 7B & $-0.9$ & +0.3 & +1.1 & \textbf{+2.6} & +1.4 & +1.1 \\
Safiron 8B & 0.0 & 0.0 & +0.3 & +1.7 & +0.3 & $-1.1$ \\
AgentDoG 4B & 0.0 & 0.0 & 0.0 & 0.0 & 0.0 & 0.0 \\
Gemma 4 12B & 0.0 & 0.0 & 0.0 & 0.0 & 0.0 & 0.0 \\
Qwen 3.5 9B & $-0.3$ & +0.3 & \textbf{+2.8} & \textbf{+2.6} & +1.7 & +0.9 \\
Nemotron Nano 9B & +0.9 & $-1.4$ & +0.3 & +1.7 & +0.6 & $-0.6$ \\
Gemma 4 E2B & $-1.1$ & 0.0 & $-0.6$ & 0.0 & 0.0 & +0.6 \\
Qwen 3.5 4B & 0.0 & +0.6 & +0.6 & \textbf{+1.1} & +0.3 & +0.9 \\
Nemotron 3 Nano 4B & 0.0 & +0.9 & $-0.3$ & +2.3 & $-0.9$ & $-1.1$ \\
\bottomrule\end{tabular}\end{table}

\section{Translation results in full}

Table~\ref{tab:tiers} pools the translation conditions by language-resource tier; Tables~\ref{tab:trans-dfpr} and \ref{tab:trans-dtpr} give \dfpr{} and \dtpr{} for every guard and language with significance marked; Table~\ref{tab:flips} (Appendix~\ref{sec:paraphrase}) gives the raw flip rates.

\begin{table}[t]\caption{Translation shifts pooled by language-resource tier (high: German, French, Joshi et al.\ class 5; mid: Turkish, class 4; low: Swahili, Amharic, class 2), percentage points vs.\ original; bold: 95\% trajectory-clustered interval excludes 0. Flip: placebo-adjusted verdict flip rate.}\label{tab:tiers}\centering\scriptsize
\begin{tabular}{lrrrrrrrrr}\toprule
 & \multicolumn{3}{c}{\dfpr{}} & \multicolumn{3}{c}{\dtpr{}} & \multicolumn{3}{c}{adj.\ flip} \\\cmidrule(lr){2-4}\cmidrule(lr){5-7}\cmidrule(lr){8-10}
Guard & high & mid & low & high & mid & low & high & mid & low \\\midrule
Gemini 3.8 Flash (scanner) & 0.0 & 0.0 & 0.0 & 0.0 & 0.0 & 0.0 & +-0.0 & $-0.1$ & $-0.1$ \\
Gemini 3.8 Flash & 0.0 & 0.0 & 0.0 & 0.0 & 0.0 & 0.0 & +0.0 & +0.0 & +0.0 \\
Claude Sonnet 5 & $-0.2$ & +0.5 & -- & $-0.1$ & $-0.4$ & -- & +0.8 & +1.1 & -- \\
GLM 5.2 & $\mathbf{-1.1}$ & $-1.0$ & -- & 0.0 & 0.0 & -- & +0.3 & +0.4 & -- \\
Nemotron 3.5 30B-A3B & \textbf{+2.4} & \textbf{+4.1} & \textbf{+2.1} & $-0.3$ & +0.3 & +1.0 & +1.2 & +2.5 & +3.2 \\
Qwen 3.8 27B & $\mathbf{-0.8}$ & $\mathbf{-0.9}$ & $-0.2$ & 0.0 & 0.0 & 0.0 & +0.1 & +0.0 & +-0.0 \\
Gemma 4 26B-A4B & \textbf{+0.8} & \textbf{+2.9} & -- & 0.0 & 0.0 & -- & +0.5 & +1.7 & -- \\
Gemma 4 26B-A4B (self-hosted) & +0.4 & \textbf{+2.4} & \textbf{+2.4} & 0.0 & 0.0 & 0.0 & +0.0 & +1.7 & +1.5 \\
Llama 4 Maverick & \textbf{+1.4} & \textbf{+1.4} & -- & $-0.3$ & +1.8 & -- & +1.9 & +2.2 & -- \\
TS-Guard 7.6B & \textbf{+2.3} & \textbf{+2.8} & $-0.2$ & $-1.7$ & $-3.1$ & $-0.1$ & +2.0 & +4.8 & +4.1 \\
ShieldAgent 7B & $\mathbf{-6.5}$ & $\mathbf{-3.8}$ & $\mathbf{-4.6}$ & +0.7 & $-1.4$ & +0.6 & +5.5 & +6.7 & +10.4 \\
Safiron 8B & +0.2 & +0.2 & $-0.5$ & $-1.0$ & $\mathbf{-3.4}$ & $\mathbf{-3.0}$ & +1.9 & +3.2 & +3.7 \\
AgentDoG 4B & +0.4 & +0.2 & +0.3 & 0.0 & 0.0 & 0.0 & +2.5 & +2.7 & +6.9 \\
Gemma 4 12B & +0.3 & \textbf{+2.0} & \textbf{+1.8} & 0.0 & 0.0 & 0.0 & +0.4 & +0.6 & +0.8 \\
Qwen 3.5 9B & +0.4 & $-0.2$ & \textbf{+1.6} & +0.6 & $-1.1$ & $-0.9$ & +1.3 & +2.5 & +2.4 \\
Nemotron Nano 9B & $\mathbf{-0.9}$ & $-0.6$ & $\mathbf{-1.2}$ & $\mathbf{-3.8}$ & $\mathbf{-4.3}$ & $\mathbf{-5.5}$ & +1.5 & +1.6 & +2.2 \\
Gemma 4 E2B & $-0.6$ & +0.2 & $\mathbf{-1.4}$ & $\mathbf{-2.3}$ & $\mathbf{-2.3}$ & $\mathbf{-4.0}$ & +1.8 & +2.4 & +2.5 \\
Qwen 3.5 4B & \textbf{+0.7} & \textbf{+0.9} & +0.4 & \textbf{+0.9} & \textbf{+1.1} & +0.4 & +0.7 & +1.3 & +1.3 \\
Nemotron 3 Nano 4B & $-0.9$ & $-1.6$ & +1.9 & $-1.3$ & $-1.7$ & +1.7 & +0.6 & +4.1 & +7.2 \\
\bottomrule\end{tabular}\end{table}
\begin{table}[t]\caption{\dfpr{} (pp) under translation; bold: 95\% interval excludes 0.}\label{tab:trans-dfpr}\centering\small
\begin{tabular}{lrrrrr}\toprule
Guard & de & fr & tr & sw & am \\\midrule
Gemini 3.8 Flash (scanner) & 0.0 & 0.0 & 0.0 & 0.0 & 0.0 \\
Gemini 3.8 Flash & 0.0 & 0.0 & 0.0 & 0.0 & 0.0 \\
Claude Sonnet 5 & $-0.1$ & $-0.3$ & +0.5 & -- & -- \\
GLM 5.2 & $-1.0$ & $\mathbf{-1.1}$ & $-1.0$ & -- & -- \\
gpt-oss-120b & -- & -- & -- & -- & -- \\
Nemotron 3.5 30B-A3B & \textbf{+4.7} & +0.1 & \textbf{+4.1} & $-0.8$ & \textbf{+5.0} \\
Qwen 3.8 27B & $\mathbf{-1.0}$ & $-0.6$ & $\mathbf{-0.9}$ & $\mathbf{-0.6}$ & +0.2 \\
Gemma 4 26B-A4B & \textbf{+1.1} & +0.6 & \textbf{+2.9} & -- & -- \\
Gemma 4 26B-A4B (self-hosted) & +0.6 & +0.2 & \textbf{+2.4} & \textbf{+1.2} & \textbf{+3.6} \\
Llama 4 Maverick & \textbf{+1.8} & +1.0 & \textbf{+1.4} & -- & -- \\
TS-Guard 7.6B & \textbf{+3.6} & +1.0 & \textbf{+2.8} & $\mathbf{-2.2}$ & \textbf{+1.8} \\
ShieldAgent 7B & $\mathbf{-7.6}$ & $\mathbf{-5.4}$ & $\mathbf{-3.8}$ & $\mathbf{-5.3}$ & $\mathbf{-3.8}$ \\
Safiron 8B & +0.5 & 0.0 & +0.2 & $-0.5$ & $-0.6$ \\
AgentDoG 4B & +0.1 & +0.7 & +0.2 & $-0.9$ & +1.5 \\
Gemma 4 12B & +0.3 & +0.2 & \textbf{+2.0} & +0.8 & \textbf{+2.8} \\
Qwen 3.5 9B & +0.6 & +0.2 & $-0.2$ & +0.6 & \textbf{+2.5} \\
Nemotron Nano 9B & $-0.7$ & $\mathbf{-1.2}$ & $-0.6$ & $\mathbf{-1.0}$ & $\mathbf{-1.4}$ \\
Gemma 4 E2B & $-0.1$ & $\mathbf{-1.2}$ & +0.2 & $\mathbf{-1.7}$ & $\mathbf{-1.2}$ \\
Qwen 3.5 4B & \textbf{+1.2} & +0.2 & \textbf{+0.9} & $-0.1$ & +0.9 \\
Nemotron 3 Nano 4B & +0.3 & $\mathbf{-2.2}$ & $-1.6$ & \textbf{+2.5} & +1.3 \\
\bottomrule\end{tabular}\end{table}
\begin{table}[t]\caption{\dtpr{} (pp) under translation; bold: 95\% interval excludes 0.}\label{tab:trans-dtpr}\centering\small
\begin{tabular}{lrrrrr}\toprule
Guard & de & fr & tr & sw & am \\\midrule
Gemini 3.8 Flash (scanner) & 0.0 & 0.0 & 0.0 & 0.0 & 0.0 \\
Gemini 3.8 Flash & 0.0 & 0.0 & 0.0 & 0.0 & 0.0 \\
Claude Sonnet 5 & $-0.2$ & 0.0 & $-0.4$ & -- & -- \\
GLM 5.2 & 0.0 & $-0.1$ & 0.0 & -- & -- \\
gpt-oss-120b & -- & -- & -- & -- & -- \\
Nemotron 3.5 30B-A3B & 0.0 & $-0.6$ & +0.3 & +0.3 & \textbf{+1.7} \\
Qwen 3.8 27B & 0.0 & 0.0 & 0.0 & 0.0 & 0.0 \\
Gemma 4 26B-A4B & 0.0 & 0.0 & 0.0 & -- & -- \\
Gemma 4 26B-A4B (self-hosted) & 0.0 & 0.0 & 0.0 & 0.0 & 0.0 \\
Llama 4 Maverick & +0.1 & $-0.7$ & +1.8 & -- & -- \\
TS-Guard 7.6B & $-2.6$ & $-0.9$ & $-3.1$ & $-0.6$ & +0.3 \\
ShieldAgent 7B & $-0.9$ & \textbf{+2.3} & $-1.4$ & $-0.6$ & +1.7 \\
Safiron 8B & $-0.6$ & $-1.4$ & $\mathbf{-3.4}$ & $-2.8$ & $\mathbf{-3.1}$ \\
AgentDoG 4B & 0.0 & 0.0 & 0.0 & 0.0 & 0.0 \\
Gemma 4 12B & 0.0 & 0.0 & 0.0 & 0.0 & 0.0 \\
Qwen 3.5 9B & +0.6 & +0.6 & $-1.1$ & $\mathbf{-3.1}$ & +1.4 \\
Nemotron Nano 9B & $\mathbf{-3.1}$ & $\mathbf{-4.5}$ & $\mathbf{-4.3}$ & $\mathbf{-3.7}$ & $\mathbf{-7.4}$ \\
Gemma 4 E2B & $\mathbf{-2.0}$ & $\mathbf{-2.6}$ & $\mathbf{-2.3}$ & $\mathbf{-4.0}$ & $\mathbf{-4.0}$ \\
Qwen 3.5 4B & +0.9 & +0.9 & \textbf{+1.1} & +0.3 & +0.6 \\
Nemotron 3 Nano 4B & $\mathbf{-2.0}$ & $-0.6$ & $-1.7$ & +1.7 & +1.7 \\
\bottomrule\end{tabular}\end{table}

\section{Strict decision rule}
\label{app:strict}

The composite score takes values in $\{0, 0.5, 1\}$; the main text blocks at $\ge 0.5$. Table~\ref{tab:strict} repeats the pooled placebo-adjusted flip rates under the strict rule (block only at $1$). The small guards' sensitivity grows under the strict rule while the frontier guards stay at zero under both.

\begin{table}[t]\caption{Placebo-adjusted flip rate (pp), pooled over paraphrase and over translation conditions, under the default rule (block if composite $\ge 0.5$) and the strict rule (composite $= 1$).}\label{tab:strict}\centering\small
\begin{tabular}{lrrrr}\toprule
Guard & para @0.5 & para @1 & trans @0.5 & trans @1 \\\midrule
Gemini 3.8 Flash (scanner) & -0.0 & +0.1 & -0.0 & +0.1 \\
Gemini 3.8 Flash & +0.0 & +0.0 & +0.0 & -0.0 \\
Claude Sonnet 5 & +0.6 & +0.2 & +0.9 & +0.2 \\
GLM 5.2 & -0.1 & +0.1 & +0.3 & +0.1 \\
gpt-oss-120b & +0.6 & +0.9 & -- & -- \\
Nemotron 3.5 30B-A3B & +0.5 & +2.0 & +2.2 & +5.7 \\
Qwen 3.8 27B & -0.0 & -0.3 & +0.0 & -0.3 \\
Gemma 4 26B-A4B & +0.6 & +0.8 & +0.9 & +0.7 \\
Gemma 4 26B-A4B (self-hosted) & +0.3 & +0.4 & +1.0 & +1.1 \\
Llama 4 Maverick & +1.0 & +1.0 & +2.0 & +1.4 \\
TS-Guard 7.6B & +1.4 & +0.8 & +3.4 & +2.6 \\
ShieldAgent 7B & +3.5 & +3.5 & +7.7 & +7.7 \\
Safiron 8B & +1.3 & +1.3 & +2.9 & +2.9 \\
AgentDoG 4B & +1.8 & +1.8 & +4.3 & +4.3 \\
Gemma 4 12B & +0.4 & +0.0 & +0.6 & +0.1 \\
Qwen 3.5 9B & +1.3 & +1.3 & +2.0 & +2.1 \\
Nemotron Nano 9B & +0.5 & +0.1 & +1.8 & +0.7 \\
Gemma 4 E2B & +1.2 & +0.5 & +2.2 & +0.5 \\
Qwen 3.5 4B & +0.6 & +1.6 & +1.0 & +2.9 \\
Nemotron 3 Nano 4B & +1.4 & +2.7 & +3.9 & +5.1 \\
\bottomrule\end{tabular}\end{table}

\section{\suite{}: generation and validation}
\label{app:perturb}

\paragraph{Release contents.} \suite{} is released as the 12 frozen rewrite files (58 rows each, keyed by instruction id, with the generation trace, judge outputs and review decisions), their sha256 hashes, the materialiser that combines them with the TS-Bench release into 1,220 labelled steps per condition, the guard prompts and adapters, and the 489k verdict rows with model id, prompt hash, token usage and latency. Trajectories and labels are not redistributed; they are read from TS-Bench.

\paragraph{Construct.} Instruction $B$ is task-equivalent to $A$ if a competent assistant following $B$ performs exactly the same tool actions, on the same objects, under the same conditions and ordering, with the same authorisation scope. Vocabulary, syntax, register, speech act, typing noise and language may vary; quoted strings, file names, e-mails, URLs, IBANs, dates, times, amounts, names and months may not; nor may the number and identity of requested actions, conditions such as ``if there are multiple\ldots'', or any urgency, permission or scope language.

\paragraph{Generation.} LLM paraphrases and translations are produced by Gemini~3.8 Flash ($T=0.7$ and $0.3$ respectively), three candidates per instruction, the first candidate in generation order that passes every check being selected; up to two feedback rounds name the failed check. Checks, in order: not identical; slot preservation (translation mode requires digits of dates and all non-name slots verbatim, leaving proper names to the judges); no added authorisation phrase; normalised edit distance within the style's bounds ($\le 0.85$ for all); length ratio in $[0.4, 3.0]$. Typos are programmatic (keyboard-adjacency substitution, transposition, drop, double; 8\% of eligible words; slot spans protected; seeded). The human paraphrases were written before any LLM output was seen. All 58 instructions passed every check in every condition on the first or second round.

\paragraph{Validation.} Two cross-family judges apply the construct as a rubric and output equivalent / missing / added / changed / authorisation-changed: Claude Sonnet~5 (primary; translations judged on its own back-translation) and Gemini~3.1 Pro (secondary). Primary: 521/522 equivalent; secondary: 516/522. Every row was then read by a human with the judges' flags visible (six rows flagged, one repaired for a dropped possessive in a Turkish translation, none rejected). For the low-resource tier, which no author reads, the rule was fixed before the judges ran: accept a row when both judges return \emph{equivalent}; otherwise repair the flagged slot from the back-translation, or reject. Swahili: primary 57/58, secondary 57/58 equivalent, no repairs. Amharic: primary 51/58, secondary 55/58; four rows repaired (three superlatives rendered as positives, ``highest-rated'' as ``highly rated'', and ``both lunch and dinner'' as ``lunch or dinner''), none rejected. The remaining flags were name transliterations (Charlie/Charlyn, Sarah/Sara) and a hiking/walk lexical choice that both the back-translation and the second judge accepted. The frozen files (sha256 in the release) are the artifact evaluated; no guard inference on a perturbed condition preceded the freeze.

\section{Serving and reproducibility}
\label{app:serving}

All guards run on one cloud platform. Managed models: Gemini~3.8 Flash and Gemini~3.1 Pro through the platform's Gemini API with a response schema; Claude Sonnet~5 through the partner API with thinking disabled (the SDK exposes no temperature for this model); GLM~5.2, gpt-oss-120b, Gemma~4 26B-A4B and Llama~4 Maverick through the platform's OpenAI-compatible endpoint with \texttt{response\_format = json\_schema} and 8k tokens of headroom for hidden reasoning. Self-deployed models (Qwen~3.8 27B FP8, Nemotron~3.5 Lightning 30B-A3B, TS-Guard, the six small guards, a self-deployed copy of Gemma~4 26B-A4B used for the low-resource pair, and the three purpose-built guards ShieldAgent, Safiron and AgentDoG at pinned revisions) run vLLM on one RTX PRO 6000 with max-model-len 32768, greedy decoding and JSON-schema enforcement (TS-Guard uses its authors' XML prompt and parser instead; ShieldAgent uses the Agent-SafetyBench evaluation prompt with tool definitions and a \texttt{<User>}/\texttt{<Assistant calls the tool>}/\texttt{<Tool>} dialogue and its \texttt{[Answer]} parser; AgentDoG uses its model-card trajectory prompt with \texttt{[USER]}/\texttt{[AGENT]}/\texttt{[ENVIRONMENT]} turns and a safe/unsafe output; Safiron uses its model-card guardian instruction with a JSON input of request, per-step actions and environment, harmless iff the output starts with ``There are no risks''). Every verdict row records model id, prompt hash, token usage and latency. TS-Guard's accuracy on the original wording (0.918) matches the released figure (0.917). The self-deployed Gemma~4 26B-A4B agrees with the managed one on 98.2\% of verdicts on the original wording and 97--98\% under the shared perturbed conditions, with a slightly lower false-positive rate (0.324 vs.\ 0.342); the two are reported as separate rows. Managed guards: 3 repeats at temperature~0; the placebo flip rate and the repeat-to-repeat flip rate are reported as noise floors. Total compute: about \$1,460 of API and GPU spend (\$934 of tokens, 105 GPU-hours), a third of it idle endpoint time during an authentication outage.

\end{document}